\documentclass[aps,pra,twocolumn,superscriptaddress,showpacs]{revtex4}

\usepackage{graphicx}

\newcommand{\ket}[1]{\mbox{$\,\mid \! #1 \, \rangle$}}
\newcommand{\bra}[1]{\mbox{$\langle \, #1 \! \mid \,$}}

\newcommand{\ketbra}[1]{\,\!\mid \! #1 \, \rangle\langle \, #1 \! \mid}

\newcommand{\Had}{\ensuremath{\mathcal{H}}}

\newcommand{\cref}[1]{Chapter~\ref{#1}}
\newcommand{\sref}[1]{Section~\ref{#1}}
\newcommand{\fref}[1]{Fig.~\ref{#1}}
\newcommand{\eref}[1]{Eq.~\ref{#1}}
\newcommand{\tref}[1]{Tab.~\ref{#1}}

\newcommand{\markproj}[1]{#1'}
\newcommand{\inimode}{s}

\newcommand{\rbox}[1]{\raisebox{1.5ex}[-1.5ex]{#1}}

\begin{document}


\title{Multi-qubit entanglement engineering via projective
measurements}


\author{Witlef Wieczorek}
\email[]{witlef.wieczorek@mpq.mpg.de}
\author{Nikolai Kiesel}
\author{Christian Schmid}
\author{Harald Weinfurter}

\affiliation{Max-Planck-Institut f{\"u}r Quantenoptik, Hans-Kopfermann-Strasse 1, D-85748 Garching, Germany}
\affiliation{Department f\"ur Physik, Ludwig-Maximilians-Universit{\"a}t, D-80797 M{\"u}nchen, Germany}


\date{\today} 

\begin{abstract}

So far, various multi-photon entangled states have been observed experimentally by using different experimental set-ups. Here, we present a scheme to realize many SLOCC-inequivalent states of three and four qubits via projective measurements on suitable entangled states. We demonstrate how these states can be observed experimentally in a single set-up and study the feasibility of the implementation with present-day technology.

\end{abstract}


\maketitle

\section{Introduction}

Entangled states are an essential resource for quantum information applications. Recently, the equivalence under stochastic local operations and classical communication (SLOCC) was successfully used to classify multi-partite entanglement \cite{Ben00,Dur00,Ver02,Lam06}. This classification is particularly relevant for evaluating the use of states for multi-party quantum communication as states of the same SLOCC class can be employed for the same applications. Therefore, the experimental realization of different SLOCC-inequivalent states is highly desirable. 

So far, several SLOCC-inequivalent states have been realized in various physical systems. The biggest variety of states was observed in experiments that rely on photonic qubits (e.g.~\cite{Bou99}). However, typical for this experimental approach is its inherent inflexibility. The design of the necessary optical network is especially tailored to the particular state that should be observed. Consequently, once a particular network is built it will not offer, in general, the choice between different SLOCC-inequivalent states. Recently, a linear optics experiment was performed that broke with this inflexibility \cite{Wie08} by allowing the observation of an entire family of SLOCC-inequivalent four-photon entangled states. Essentially, this was achieved by multi-photon interference. 

Here we show that projective measurements on subsystems can provide another means of preparing SLOCC-inequivalent classes of entangled states. It is well known that atomic entangled states, even from different SLOCC-classes, can be remotely prepared by projective measurements on photons \cite{Cab99,Bas07qph}, which previously have been entangled with the atoms, i.e., by a measurement of typically half of the total multi-partite entangled state. Here, in contrast, we focus on the property of certain symmetric multi-partite entangled states that allow a more flexible preparation of families of SLOCC-inequivalent types of entanglement by projective measurements on small subsystems. The initial $n$-qubit symmetric states can be observed in linear optics set-ups that distribute $n$ photons of a single spatial mode to $n$ different output modes. Subsequent projective measurements on these $n$-qubit states will yield states belonging to different SLOCC classes. We focus in the following on the case of $n=4$ and $n=5$ and demonstrate that our approach can be realized by using a single linear optical set-up only.

The paper is structured as follows. In \sref{sec-1} we discuss the effect of projective measurements on particular symmetric states. We begin our investigations with SLOCC-inequivalent three-qubit states obtained from the four-qubit symmetric Dicke state with two excitations $\ket{D_4^{(2)}}$ \cite{Dic54,Kie07,Sch07ProcNato}.
Further, we show how to obtain SLOCC-inequivalent four-qubit entangled states like, e.g., the states $\ket{GHZ_4}$, $\ket{W_4}$ and even $\ket{D_4^{(2)}}$, via projective measurements on five-qubit states, which are given by superpositions of two symmetric Dicke states \cite{ScP07}. In  \sref{sec-2} we will discuss the experimental implementation of the proposed schemes. We recapitulate the experiment of Ref.~\cite{Kie07} that lead to the observation of $\ket{D_4^{(2)}}$ and discuss the feasibility of an extension in order to observe the five-qubit states.

\section{\label{sec-1}Projective measurements on particular symmetric states}

In their seminal work D\"ur \emph{et al.} \cite{Dur00} discovered that only two SLOCC-inequivalent classes of genuine tri-partite entanglement exist: the GHZ and W class. Well known representatives of these classes are the states $\ket{GHZ_3}=1/\sqrt{2}(\ket{HHH}+\ket{VVV})$ and $\ket{W_3}=1/\sqrt{3}(\ket{HHV}+\ket{HVH}+\ket{VHH})$, respectively. We utilize the notation for polarization encoded qubits throughout this work, e.g.~$\vert HHV\rangle=\vert H\rangle_a\otimes\vert H\rangle_b\otimes\vert V\rangle_c$ and $\vert H\rangle$ or $\vert V\rangle$ mean linear horizontal ($H$) or vertical ($V$) polarization of photons, respectively, and the subscript denotes the spatial mode of each photon. In contrast to the three qubit case, the SLOCC classification of four-partite entangled states is much richer, containing infinitely many SLOCC-inequivalent four-partite entangled states \cite{Ver02,Lam07}. 

In the following we show that via projective measurements on particular symmetric states, SLOCC-inequivalent entangled states of a lower qubit number can be obtained. To this end, we consider particular members of the family of symmetric Dicke states \cite{Dic54}. Generally, a symmetric $N$-qubit Dicke state with $m$ excitations, denoted as $\ket{D_N^{(m)}}$, is, again in the notation of polarization encoded photonic qubits, the equally weighted superposition of all possible permutations of $N$-qubit product states with $m$ vertically and $N-m$ horizontally polarized photons. 

\subsection{Projections of the four-qubit Dicke state $\ket{D_4^{(2)}}$}

First, we aim at obtaining states from the two inequivalent tri-partite entanglement classes by applying projective measurements on a four-qubit entangled state. The symmetric Dicke state $\ket{D_4^{(2)}}=\frac{1}{\sqrt{6}}(\vert HHVV\rangle+\vert HVHV\rangle+\vert VHHV\rangle+\vert HVVH\rangle+\vert VHVH\rangle+\vert VVHH\rangle)$ turned out to be useful for this purpose \cite{Kie07}. Here, we will analyze in more detail which three-qubit states can be obtained. 

Generally, an arbitrary projective measurement can be expressed by $P(\markproj{\alpha},\markproj{\epsilon}):=\ket{\markproj{\alpha},\markproj{\epsilon}}\bra{\markproj{\alpha},\markproj{\epsilon}}$ with $\ket{\markproj{\alpha},\markproj{\epsilon}}=\markproj{\alpha} \ket{H}+\markproj{\beta}
e^{i\markproj{\epsilon}}\ket{V}$ (all parameters real and $\markproj{\alpha}^2+\markproj{\beta}^2=1$). The projection $P(\markproj{\alpha},\markproj{\epsilon})$ applied on $\ket{D_4^{(2)}}$ leads to the three-qubit states
\begin{equation}
\label{eq-3qubitDicke}
\propto\markproj{\alpha}\ket{D_3^{(2)}}+\markproj{\beta}e^{-i\markproj{\epsilon}}\ket{D_3^{(1)}},
\end{equation}
which are arbitrary superpositions of the two entangled, symmetric three-qubit Dicke states (\tref{tab-3qubit}).

\begin{table}
\caption{\label{tab-3qubit} Three-qubit states obtained by a single projective measurement on the state $\ket{D_4^{(2)}}$, cp.~\eref{eq-3qubitDicke}.}
\begin{ruledtabular}
\begin{tabular}{l|l|l||c}
$\markproj{\alpha}$ & $\markproj{\beta}$ & $\markproj{\epsilon}$ & state\\\hline\hline
$\cos{\markproj{\theta}}$ & $\sin{\markproj{\theta}}$ & $\markproj{\epsilon}$ & $\cos{\markproj{\theta}}\ket{D_3^{(2)}}+e^{-i\markproj{\epsilon}}\sin{\markproj{\theta}}\ket{D_3^{(1)}}$\\
$1$ & $0$ & $-$ & $\ket{D_3^{(2)}}\equiv\ket{\overline{W}_3}$ \\
$0$ & $1$ & $-$ & $\ket{D_3^{(1)}}\equiv\ket{W_3}$ \\
$\frac{1}{\sqrt{2}}$ & $\frac{1}{\sqrt{2}}$ & $0$ & $(\ket{D_3^{(1)}}+\ket{D_3^{(2)}})/\sqrt{2}\equiv\ket{G_3^+}$ \\
$\frac{1}{\sqrt{2}}$ & $\frac{1}{\sqrt{2}}$ & $\pi$ & $(\ket{D_3^{(1)}}-\ket{D_3^{(2)}})/\sqrt{2}\equiv\ket{G_3^-}$ \\
\end{tabular}
\end{ruledtabular}
\end{table}

\begin{figure}[thp]
\includegraphics{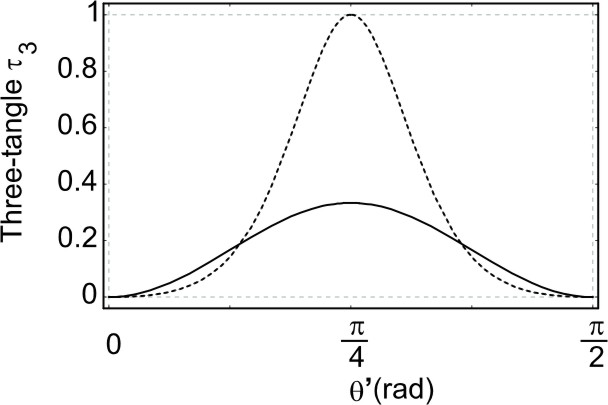}
\caption{\label{fig-tangle} Three-tangle $\tau_3$ for the states of \eref{eq-3qubitDicke} without (straight) and with (dashed) application of the transformation $\mathcal{T}_{+}\otimes\mathcal{T}_{+}\otimes\mathcal{T}_{+}$ (where $\markproj{\alpha}=\cos{\markproj{\theta}}$ and $\markproj{\epsilon}=0$).}
\end{figure}

To analyze the entanglement of the states we choose as a suitable entanglement measure the three-tangle $\tau_3$ \cite{Cof00}, which distinguishes the W and GHZ class as only for $GHZ$ type entangled states $\tau_3$ is non-zero \cite{Dur00}. The solid line in \fref{fig-tangle} shows $\tau_3$ for the states of \eref{eq-3qubitDicke} in dependence of $\markproj{\theta}$ ($\markproj{\alpha}=\cos{\markproj{\theta}}$). It is found that the three-tangle is zero for $\markproj{\theta}=0,\pi/2$ ($\markproj{\alpha}=0,1$), which corresponds to a measurement in the computational basis. There, we obtain states from the W class, namely $\ket{W_3}\equiv\ket{D_3^{(1)}}$ and $\ket{\overline{W}_3}=1/\sqrt{3}(\ket{VVH}+\ket{VHV}+\ket{HVV})\equiv\ket{D_3^{(2)}}$, respectively. For all other values of $\markproj{\theta}$, $\tau_3\neq0$ implying that these states belong to the GHZ class. The maximal value of $\tau_3=1/3$ is obtained for $\markproj{\theta}=\pi/4$ corresponding to a measurement in the $(\pm)$-basis, where $\ket{\pm}=1/\sqrt{2}(\ket{H}\pm\ket{V})$. For $\markproj{\theta}=\pi/4$ and $\markproj{\epsilon}=0,\pi$ one obtains the $G_3$ states \cite{Sen03a} with $\ket{G_3^+}=\frac{1}{\sqrt{2}}(\ket{\overline{W}_3}+\ket{W_3})$ and $\ket{G_3^-}=\frac{1}{\sqrt{2}}(\ket{\overline{W}_3}-\ket{W_3})$, respectively. 

The $G_3$ states can be transformed directly into the $GHZ_3$ state, which has the maximal possible three-tangle $\tau_3=1$, via the stochastic local operations (local filtering)
\begin{eqnarray*}
\mathcal{T}_{+}&=&\Had\left[\frac{1}{2}\left((\frac{1}{\sqrt{3}}+i)\cdot\openone +(\frac{1}{\sqrt{3}}-i)\cdot\sigma_z\right)\right]\Had,\\
\mathcal{T}_{-}&=&\Had\left[\frac{1}{2}\left((\frac{1}{\sqrt{3}}+i)\cdot\sigma_x +i(\frac{1}{\sqrt{3}}-i)\cdot\sigma_y\right)\right]\Had,
\end{eqnarray*}
where $\sigma_x$, $\sigma_y$ and $\sigma_z$ are the Pauli spin matrices and $\Had$ is the Hadamard transformation, in the following way:
\begin{eqnarray}
\nonumber\left(\mathcal{T}_{+}\otimes\mathcal{T}_{+}\otimes\mathcal{T}_{+}\right)\ket{G_3^+}&=&\frac{1}{3}\ket{GHZ_3},\\
\label{eq-trafosGHZ}\left(\mathcal{T}_{-}\otimes\mathcal{T}_{-}\otimes\mathcal{T}_{-}\right)\ket{G_3^-}&=&\frac{1}{3}\ket{GHZ_3}.
\end{eqnarray}
Though these operations perform the desired transformation, they only do so with a success probability of $1/9$. \fref{fig-tangle} shows the three-tangle when the operation $\mathcal{T}_{+}\otimes\mathcal{T}_{+}\otimes\mathcal{T}_{+}$ is applied successfully to all states of \eref{eq-3qubitDicke}. For $\markproj{\theta}=\pi/4$ the three-tangle is indeed increased to its maximal value of $\tau_3=1$. 


\subsection{Projections of superpositions of five-qubit Dicke states\label{sec-TDt5}}

Extending the idea described before, SLOCC-inequivalent four-partite entangled states can be obtained from suitable five-qubit symmetric states via projective measurements. Here we consider an arbitrary superposition of the two symmetric five-qubit Dicke states $\ket{D_5^{(2)}}$ and $\ket{D_5^{(3)}}$:
\begin{equation}
\label{eq-5photres}
\ket{\Delta_5}=\alpha \ket{D_5^{(2)}}+\beta e^{i\epsilon}\ket{D_5^{(3)}}.
\end{equation}
We note that these states can also be seen as a natural choice as they are obtained via a single projective measurement on the six-qubit Dicke state $\ket{D_6^{(3)}}$. The states $\ket{\Delta_5}$ belong to two different SLOCC classes. The first class occurs for $\alpha=0$ or $\beta=0$, as the states $\ket{D_5^{(2)}}$ and $\ket{D_5^{(3)}}$ can be transformed into each other by spin-flipping all qubits. The second class is given for $\alpha\neq0$ and $\beta\neq0$. The weighting and phase between the terms of $\ket{\Delta_5}$ can be changed easily via the SLOCC-operations $\mathcal{T}_r^{\otimes 5}=\mathcal{T}_r\otimes\mathcal{T}_r\otimes\mathcal{T}_r\otimes\mathcal{T}_r\otimes\mathcal{T}_r$ with $\mathcal{T}_r=[(1+1/r)\openone+(1-1/r)\sigma_z]/2$ and $r\neq0$ complex. To obtain a new ratio of parameters $\overline{\alpha}/(\overline{\beta} e^{i\overline{\epsilon}})$, $r$ needs to be chosen as $\beta\overline{\alpha}e^{i\epsilon}/(\overline{\beta}\alpha e^{i\overline{\epsilon}})$. Note that this reasoning could also be applied for the states of \eref{eq-3qubitDicke}.

A single projective measurement $P(\markproj{\alpha},\markproj{\epsilon})$ applied on $\ket{\Delta_5}$ yields the four-qubit entangled states 
\begin{eqnarray}
\label{eq-4photext}
\nonumber\ket{\Delta_4}&\propto&\alpha\markproj{\beta} e^{-i\markproj{\epsilon}}\ket{D_4^{(1)}}+\markproj{\alpha} \beta e^{i\epsilon} \ket{D_4^{(3)}}\\
&&+(\alpha\markproj{\alpha} + \beta\markproj{\beta} e^{i (\epsilon-\markproj{\epsilon})})\sqrt{6/4}\ket{D_4^{(2)}}.
\end{eqnarray}
These states are superpositions of all symmetric four-qubit entangled Dicke states. In particular, these superpositions contain the SLOCC-inequivalent family of states $\mu(\alpha,\markproj{\alpha},\epsilon,\markproj{\epsilon})\ket{GHZ_4}+\nu(\alpha,\markproj{\alpha},\epsilon,\markproj{\epsilon})\ket{D_4^{(2)}}$ (for details see \cite{StatsTrafo}), which forms according to the SLOCC-classification by Verstraete {\emph{et al.}}~\cite{Ver02} a subset of the four-qubit entangled generic family $G_{abcd}$. In the following we discuss prominent SLOCC-inequivalent states of the family \eref{eq-4photext} (see also \tref{tab-4qubit}). 

Remarkably, we obtain a four-qubit $GHZ_4$ state. This can be easily seen when we consider the state $\ket{GHZ_4^-}=1/\sqrt{2}(\ket{HHHH}-\ket{VVVV})$ under a Hadamard transformation $\mathcal{H}$ acting on each qubit:
\begin{eqnarray*}
\ket{GHZ_4^+}&\equiv&(\mathcal{H}\otimes\mathcal{H}\otimes\mathcal{H}\otimes\mathcal{H})\ket{GHZ_4^-}\\
&=&\frac{1}{\sqrt{2}}(\ket{D_4^{(1)}}+\ket{D_4^{(3)}}).
\end{eqnarray*}
We get $\ket{GHZ_4^+}$ when the amplitude of $\ket{D_4^{(2)}}$ is zero and the two remaining terms in \eref{eq-4photext} are equally balanced, i.e.~the conditions (i) $\alpha\markproj{\alpha} =-\beta\markproj{\beta} e^{i (\epsilon-\markproj{\epsilon})}$ and (ii) $\alpha\markproj{\beta} e^{-i\markproj{\epsilon}}=\markproj{\alpha} \beta e^{i\epsilon}$ are fulfilled. This holds only for $\alpha=\beta=\markproj{\alpha}=\markproj{\beta}=1/\sqrt{2}$ and $\epsilon=-\markproj{\epsilon}=\pi/2$ or $-\epsilon=\markproj{\epsilon}=\pi/2$. When we impose only condition (i), i.e.~the amplitude of $\ket{D_4^{(2)}}$ is zero [which holds for $\alpha=\sqrt{1-\markproj{\alpha}^2}$ and $\Delta\epsilon=\epsilon-\markproj{\epsilon}=(2n+1)\pi$ for $n\in\{0,1,2,...\}$], a continuous transition between the states $\ket{W_4}$ ($\alpha=1$), $\ket{GHZ_4^+}$ ($\alpha=\sqrt{1/2}$ and $\Delta\epsilon=\pi$) and $\ket{\overline{W}_4}$ ($\alpha=0$) can be achieved. 

\begin{table}
\caption{\label{tab-4qubit} Four-qubit entangled states obtained from the states $\ket{\Delta_5}$ by a single projective measurement, cp.~\eref{eq-4photext}. In analogy to \eref{eq-5photres} the states $\cos{^2\theta}\ket{D_4^{(1)}}-\sin{^2\theta}e^{2i\epsilon}\ket{D_4^{(3)}}$ belong to two SLOCC classes given by (i) $\theta=0$ or $\theta=\pi/2$ and (ii) $\theta\in(0,\pi/2)$ with $\epsilon$ arbitrary.}
\begin{ruledtabular}
\begin{tabular}{l|l|l||l|l|l||c}
$\alpha$ & $\beta$ & $\epsilon$  & $\markproj{\alpha}$ & $\markproj{\beta}$ & $\markproj{\epsilon}$ & state\\\hline\hline
$1$ & $0$ & $-$ & $1$ & $0$ & $-$ & $\ket{D_4^{(2)}}$ \\
$0$ & $1$ & $-$ & $0$ & $1$ & $-$ & $\ket{D_4^{(2)}}$ \\
$1$ & $0$ & $-$ & $0$ & $1$ & $-$ & $\ket{W_4}\equiv\ket{D_4^{(1)}}$ \\
$0$ & $1$ & $-$ & $1$ & $0$ & $-$ & $\ket{\overline{W}_4}\equiv\ket{D_4^{(3)}}$ \\
$\frac{1}{\sqrt{2}}$ & $\frac{1}{\sqrt{2}}$ & $\frac{\pi}{2}$ & $\frac{1}{\sqrt{2}}$ & $\frac{1}{\sqrt{2}}$ & $-\frac{\pi}{2}$ & $\ket{GHZ_4^+}$ \\
&&&&&& $\cos{^2\theta}\ket{D_4^{(1)}}$\\
\rbox{$\cos{\theta}$} & \rbox{$\sin{\theta}$} & \rbox{$\epsilon$} & \rbox{$\sin{\theta}$} & \rbox{$\cos{\theta}$} & \rbox{$\epsilon-\pi$} & $-\sin{^2\theta}e^{2i\epsilon}\ket{D_4^{(3)}}$ \\
\end{tabular}
\end{ruledtabular}
\end{table}

Further, three-qubit states are obtained by performing a projective measurement $P(\markproj{\markproj{\alpha}},\markproj{\markproj{\epsilon}})$ on $\ket{\Delta_4}$:
\begin{eqnarray}
\label{eq-threequbitDelta5}
\nonumber\ket{\Delta_3}&\propto&\alpha\markproj{\beta}\markproj{\markproj{\beta}} e^{-i(\markproj{\epsilon}+\markproj{\markproj{\epsilon}})}\ket{D_3^{(0)}}
+\markproj{\alpha} \beta \markproj{\markproj{\alpha}} e^{i\epsilon} \ket{D_3^{(3)}}\\
\nonumber&&+[(\alpha\markproj{\beta}\markproj{\markproj{\alpha}}e^{-i\markproj{\epsilon}})\\
\nonumber&&\hspace{2mm}+\sqrt{\frac{6}{4}}(\alpha\markproj{\alpha} + \beta\markproj{\beta} e^{i(\epsilon-\markproj{\epsilon})})\markproj{\markproj{\beta}}e^{-i\markproj{\markproj{\epsilon}}}]
\sqrt{3}\ket{D_3^{(1)}}\\
\nonumber&&+[(\markproj{\alpha}\beta\markproj{\markproj{\beta}}e^{i(\epsilon-\markproj{\markproj{\epsilon}})})\\
&&\hspace{2mm}+\sqrt{\frac{6}{4}}(\alpha\markproj{\alpha} + \beta\markproj{\beta} e^{i(\epsilon-\markproj{\epsilon})})\markproj{\markproj{\alpha}}]
\sqrt{3}\ket{D_3^{(2)}}.
\end{eqnarray}
These are {\textit{all}} permutation symmetric three-qubit states \cite{Bas08tbd}. In particular, we note that a $GHZ_3$ state can be obtained directly without the need for local operations. To show this, we consider
$\ket{GHZ_3^-}=1/\sqrt{2}(\ket{HHH}-\ket{VVV})$ under a Hadamard transformation $\mathcal{H}$ acting on each qubit:
\begin{eqnarray*}
\ket{GHZ_3^+}&\equiv&(\mathcal{H}\otimes\mathcal{H}\otimes\mathcal{H})\ket{GHZ_3^-}\\
&=&\sqrt{\frac{3}{4}}\ket{D_3^{(1)}}+\sqrt{\frac{1}{4}}\ket{D_3^{(3)}}.
\end{eqnarray*}
The state $\ket{GHZ_3^+}$ is obtained for $\alpha=\beta=\markproj{\alpha}=\markproj{\beta}=1/\sqrt{2}$, $\markproj{\markproj{\alpha}}=1$ and $\epsilon=-\markproj{\epsilon}=\pi/2$.


\section{\label{sec-2}Experimental implementation}

The states $\ket{D_4^{(2)}}$ and $\ket{\Delta_5}$ are permutation symmetric. Hence, their experimental implementation can be achieved via a symmetric distribution of photons. For the observation of the state $\ket{D_4^{(2)}}$ the necessary four photons originate from the second order emission of a collinear, type II spontaneous parametric down conversion (SPDC). For observing the states $\ket{\Delta_5}$ we will consider different experimental implementations, which are extensions of the $\ket{D_4^{(2)}}$ set-up.

\subsection{\label{sec-2a}The Dicke state $\ket{D_4^{(2)}}$}

\begin{figure}[thp]
\includegraphics{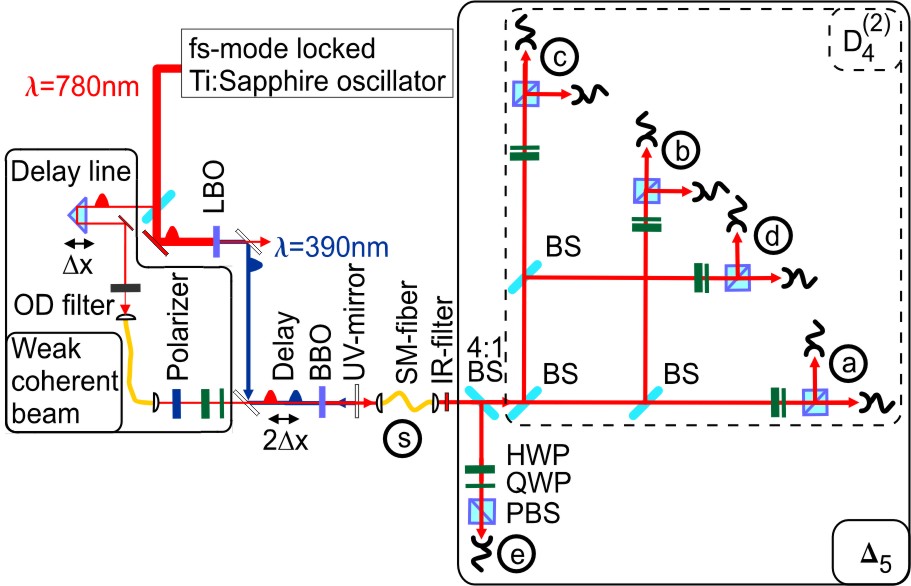}
\caption{\label{fig-setup} (Color online) Schematic experimental set-up for the observation of the state $\ket{D_4^{(2)}}$ (innermost section, dashed frame). The weak coherent beam, the 4:1 beam splitter as well as the detection in an additional mode $e$ are needed for the proposed experimental implementation of the five photon states $\ket{\Delta_5}$ (details see text).}
\end{figure}

\fref{fig-setup} shows one of the possible set-ups for the states $\ket{\Delta_5}$. As can be seen, the set-up for observing the state $\ket{D_4^{(2)}}$ is at the core of it. The state $\ket{D_4^{(2)}}$ was observed after a symmetric distribution of two horizontally and two vertically polarized photons, initially in a single spatial mode $\inimode$, onto four spatial modes ($a,b,c,d$) via three polarization-independent beam splitters (BS). The photons originate from a $\beta$-Barium borate (BBO) crystal in a type II, collinear SPDC process, which emits the state \cite{Kok00,Wei01}
\begin{equation}
\label{eq-SPDC}
\ket{\Psi_{\mathrm{dc}}}=\sqrt{1-|z_{\mathrm{dc}}|^2}\sum_{n=0}^\infty\frac{(iz_{\mathrm{dc}})^n}{n!}({\inimode}_H^\dagger {\inimode}_V^\dagger)^n\ket{\mathrm{vac}},
\end{equation}
where ${\inimode_i}^\dagger$ is the creation operator for a photon in mode $\inimode$ having polarization $i\in\{H,V\}$, $\ket{\rm{vac}}$ is the vacuum state, $z_{\mathrm{dc}}=|z_{\mathrm{dc}}|e^{i2\phi_{\mathrm{dc}}}$ with $|z_{\mathrm{dc}}|=\tanh{\tau}$ and $\tau$ depends on the pump amplitude and the coupling between the electromagnetic field and the crystal ($\tau\ll 1$). The probability to create a single pair is ${(1-|z_{\mathrm{dc}}|^2)}|z_{\mathrm{dc}}|^2$. Here, we are interested in the second order emission $\propto ({\inimode}_H^\dagger {\inimode}_V^\dagger)^2\ket{\rm{vac}}$. 

The BBO crystal was pumped by a frequency-doubled, femtosecond, mode-locked Ti:Sapphire laser. The spatial mode $\inimode$ of the photons is defined by coupling into a single mode (SM) fiber. The photons pass an interference filter (IR) reducing their spectral distinguishability. The polarization state of each photon is analyzed via a polarizing beam splitter (PBS) preceded by a half-wave- (HWP) and quarter-wave plate (QWP). Finally, the photons are detected by fiber-coupled single photon detectors. The experimental state was observed under the condition of detecting a photon in each of the four spatial modes $(a,b,c,d)$.

We found a fidelity of $F_{\mathrm{exp}}=\bra{D_4^{(2)}}\rho_{\mathrm{exp}}\ket{D_4^{(2)}}=0.844\pm0.008$ to the state $\ket{D_4^{(2)}}$. Using a generic entanglement witness $\mathcal{W}$ \cite{Bou04}, where we use the shorthand notation $\mathcal{W}(\Psi,\alpha)=\alpha\openone-\ketbra{\Psi}$ with $\alpha=\frac{2}{3}$ and $\ket{\Psi}=\ket{D_4^{(2)}}$, genuine four-partite entanglement of $\rho_{\mathrm{exp}}$ was verified: $\mathrm{Tr}[\mathcal{W}(D_4^{(2)},\frac{2}{3})\rho_{\mathrm{exp}}]=\frac{2}{3}-F_{\mathrm{exp}}=-0.177\pm0.008$. A value $<0$ is sufficient to prove genuine four-partite entanglement \cite{Bou04}. Further, by using the state-discrimination method described in \cite{Sch08DS} we were able to exclude W- and Cluster-type entanglement for the experimentally observed state.

\begin{figure}[thp]
\includegraphics{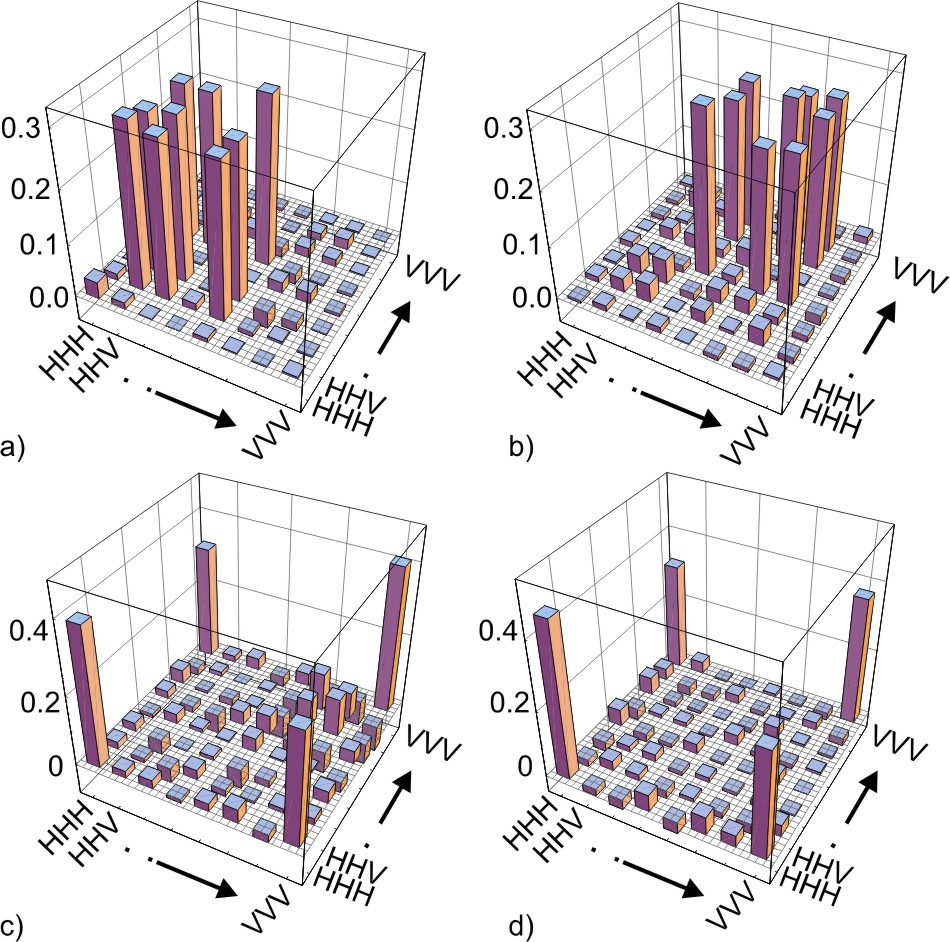}
\caption{\label{fig-states3} (Color online) Experimental density matrices for the measured $W_3$ (a) and $\overline{W}_3$ (b) states. The density matrices for the $GHZ_3$ states of (c) and (d) are calculated from the measured $G_3^+$ and $G_3^-$ states. Displayed is the real part of the corresponding density matrix.}
\end{figure}

For demonstrating that we can experimentally access states from both inequivalent tri-partite entanglement classes we performed a full state tomography to reconstruct the density matrices of the respective states. A projection measurement of the photon in mode $d$ in the computational basis yields the $W_3$ states characterized by the density matrices shown in \fref{fig-states3}(a) and (b). We calculated fidelities of $0.882\pm0.015$ and $0.835\pm0.015$ to the theoretical states $\ket{W_3}$ and $\ket{\overline{W}_3}$, respectively. Their genuine tri-partite entanglement is verified via the entanglement witnesses $\mathcal{W}(W_3,\frac{2}{3})$ and $\mathcal{W}(\overline{W}_3,\frac{2}{3})$ \cite{Bou04}, where we determined values of $-0.215\pm0.015$ and $-0.168\pm0.015$, respectively.

A measurement in the $(\pm)$-basis yields $G_3$ states, which belong to the GHZ class. If we apply the corresponding transformations (see \eref{eq-trafosGHZ}) on the measured density matrices we indeed obtain $GHZ_3$ states, see \fref{fig-states3}(c) and (d). We determined fidelities of $0.719\pm0.022$ and $0.733\pm0.024$ to a $GHZ_3$ state, respectively. An entanglement witness detecting genuine tri-partite entanglement of these states is $\mathcal{W}(GHZ_3,\frac{1}{2})$ \cite{Bou04}. We find the negative values of $-0.219\pm0.022$ and $-0.233\pm0.024$, respectively. A witness that further excludes W type entanglement is given by $\mathcal{W}(GHZ_3,\frac{3}{4})$ \cite{Aci01}. The transformed $GHZ_3$ states do not fulfill the witness's entanglement condition. However, by applying local filtering operations on this witness \cite{Haf05} we obtain values of $-0.033\pm0.026$ and $-0.029\pm0.023$, respectively, which finally proves GHZ-type entanglement with a significance of one standard deviation. 

\subsection{Towards $\ket{\Delta_5}$}

\subsubsection{Implementation}

For observing the states $\ket{\Delta_5}$ different implementations are possible. One possibility is given by the application of a projective measurement on the state $\ket{D_6^{(3)}}$ (see \sref{sec-TDt5}), where the necessary six photons originate from the third order SPDC emission. However, when implementing the state $\ket{\Delta_5}$ directly only five photons are necessary and, thus, a higher count rate should be possible. These five photons can be obtained by superimposing the four photons from the second order SPDC emission with an additional photon. The polarization of the additional photon determines the parameters $\alpha$,$\beta$ and $\epsilon$ in \eref{eq-5photres}. In the ideal case, the additional photon is obtained from a single photon source (see e.g.~\cite{Oxb05,LoS05}) that acts on demand and matches the SPDC photons spectrally, temporally and spatially. However, to our knowledge, no such source exists. Alternatively, an heralded SPDC source \cite{Roh05} can be employed, which results in practice in low count rates, since again six photons have to be detected in total. Instead, we investigate whether the single photon source can be substituted by a weak coherent beam (WCB) \cite{Rar99}, i.e., whether this simplification influences the state quality. 

This implementation is based on the set-up used for observing the state $\ket{D_4^{(2)}}$ described in \sref{sec-2a} (\fref{fig-setup}). The WCB can be derived via a beam splitter [and additional attenuation via optical density (OD) filters] from the Ti:Sapphire laser, which also pumps the BBO crystal for the SPDC process after a frequency doubling stage. The polarization of the WCB can be set arbitrarily via a polarizer, followed by a HWP and QWP. These settings determine the parameters $\alpha$,$\beta$ and $\epsilon$ in \eref{eq-5photres}. A delay line in the WCB allows to adjust the temporal overlap with the SPDC emission. The photons of both sources are overlapped collinearly in the BBO crystal and coupled in the same single mode fiber. They are symmetrically distributed onto the modes $(a,b,c,d,e)$ via a beam splitter with 4:1 splitting ratio and further splitting as described in \sref{sec-2a}.

\subsubsection{Weak coherent beam: effects}

The state of the WCB that substitutes the single photon source is \cite{Man95}
\begin{equation}
\label{eq-wcb}
\ket{\Psi_{\mathrm{w}}}=e^{-{|z_{\mathrm{w}}|}^2/2}\cdot\sum_{n=0}^\infty\frac{{(z_{\mathrm{w}})}^n}{n!}(w_j^\dagger)^n\ket{\mathrm{vac}},
\end{equation}
where $w_j^\dagger$ is the creation operator of a photon with polarization $j$ in mode $w$, $z_{\mathrm{w}}=|z_{\mathrm{w}}|e^{i\phi_{\mathrm{w}}}$, $|z_{\mathrm{w}}|^2$ is the mean photon number and $|z_{\mathrm{w}}|^2e^{-{|z_{\mathrm{w}}|}^2}$ is the probability for the photon state $\ket{1}$. The photons of the WCB and the SPDC originate from the same laser, i.e., the Ti:Sapphire laser, but travel different paths before they are coupled into the same single mode fiber. As only their relative phase is relevant, we set for the following considerations $\phi_{\mathrm{dc}}=0$, without loss of generality. In the experiment the relative phase fluctuates without an active stabilization of the relative delay of the WCB and SPDC photons. Further, the WCB compared to a real single photon source exhibits higher order terms resulting in multiple photons per pulse. We note that the phase dependence and the higher order terms have a much smaller influence if an heralded source is employed, of course, with the disadvantage of requiring, effectively, a six-photon down conversion experiment.

\begin{figure}[thp]
\includegraphics{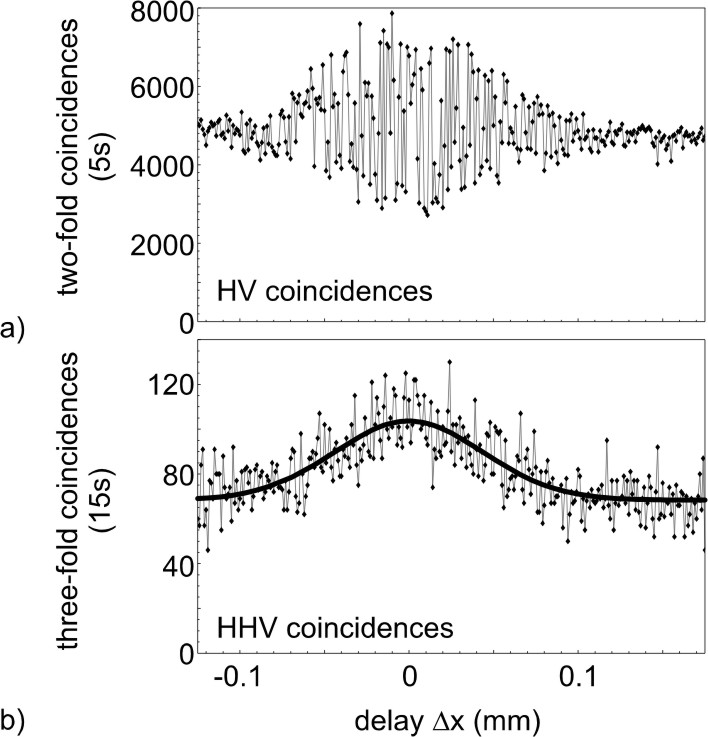}
\caption{\label{fig-inter} Coherence between the weak coherent beam and the SPDC photons. (a) Interference of the two possibilities how to obtain an $HV$ coincidence. (b) Enhanced emission (cloning) due to bosonic nature of photons. The solid line shows a Gaussian fit to the data points giving an enhancement of $1.52\pm0.03$.}
\end{figure}

We will now demonstrate effects caused by using the WCB. First, we observe quantum interference, which occurs when there are at least two indistinguishable possibilities that lead to the same detection event. In our case this becomes already observable when we consider only two photons. There, the following two possibilities exist: Two photons originate either from the SPDC emission \emph{or} the WCB. To show the interference let us assume a left circularly polarized WCB, whose two-photon term is
\begin{eqnarray}
\label{eq-wcbtwo}
&\propto& e^{2i\phi_{\mathrm{w}}}\left(w_L^{\dagger}\right)^2=e^{2i\phi_{\mathrm{w}}}\left(w_H^{\dagger}-iw_V^{\dagger}\right)^2/2\nonumber\\
&=&e^{2i\phi_{\mathrm{w}}}\left[\left(w_H^{\dagger}\right)^2-\left(w_V^{\dagger}\right)^2-2i\left(w_H^{\dagger}w_V^{\dagger}\right)\right]/2.
\end{eqnarray}
For a coherent overlap, i.e., $w_j^{\dagger}\rightarrow {\inimode}_j^{\dagger}$, the last term of \eref{eq-wcbtwo} is identical with the first order SPDC emission ($\propto {\inimode}_H^{\dagger}{\inimode}_V^{\dagger}$). Hence, for the two-fold coincidence detection event $HV$, both possibilities contribute and interfere in dependence on $\phi_{\mathrm{w}}$. This is shown in \fref{fig-inter}(a). When we change the path difference between the photons of both sources we observe an oscillation in the coincidence count rate on the order of the wavelength ($<\mu$m), which is due to the change of $\phi_{\mathrm{w}}$. The exact modulation is unresolved as $\phi_{\mathrm{w}}$ was not actively stabilized. The width of the envelope of that interference is on the order of the coherence length of the photons ($\approx100\mu$m). It indicates the spatial region for which the mode overlap is different from zero.


Furthermore, we can observe bosonic enhancement (cloning \cite{Lam02,Sci05,Mar05}), i.e., stimulation of the SPDC emission, which appears independent of the employed single photon source and enhances the total count rate. This enhancement is visible for, e.g., a horizontally polarized WCB as input and registration of a three-fold coincidence of $HHV$, \fref{fig-inter}(b). The single photon term of the WCB ($\propto w_H^{\dagger}$) and the first order emission of the SPDC ($\propto {\inimode}_H^{\dagger}{\inimode}_V^{\dagger}$) lead incoherently overlapped to $\propto w_H^{\dagger}{\inimode}_H^{\dagger}{\inimode}_V^{\dagger}\ket{\rm{vac}}=\ket{H}_w\ket{HV}_{\inimode}$. In contrast, a coherent superposition yields $\propto ({\inimode}_H^{\dagger})^2{\inimode}_V^{\dagger}\ket{\rm{vac}}=\sqrt{2}\ket{HHV}_{\inimode}$, hence, an increase by a factor of two in the count rate due to the bosonic enhancement \cite{Sun07}. This effect occurs on the order of the coherence length of the photons ($\approx100\mu$m). In the experiment we observe an enhancement of $1.52\pm0.03$. We attribute the deviation from the expected value of two to higher order emissions of the WCB and the SPDC, which add an offset to the three photon count rate.


\subsubsection{Fidelity of the states $\ket{GHZ_4^+}$ and $\ket{W_4}$}

In the following we give a quantitative estimate of the influence on the quality of the desired four-photon states when a WCB is used instead of a single photon source. To this end two effects have to be considered leading to the observation of imperfect states. Firstly, the coherent superposition of different emission orders leads, in dependence on $\phi_{\mathrm{w}}$, to the observation of a different {\emph{pure}} state. Secondly, higher order emissions cause an {\emph{admixture}} of correlated noise. The first effect can be analyzed when considering all terms from \eref{eq-SPDC} and \eref{eq-wcb} that contribute directly to five photons (yielding the state on which a projective measurement is applied):
\begin{eqnarray}
\label{eq-5phot}
\propto&-&|z_{\mathrm{dc}}|^2|z_{\mathrm{w}}|e^{i\phi_{\mathrm{w}}}w_j^\dagger ({\inimode}_H^\dagger {\inimode}_V^\dagger)^2/2\nonumber\\
&+&i|z_{\mathrm{dc}}||z_{\mathrm{w}}|^3e^{i3\phi_{\mathrm{w}}}{w_j^\dagger}^3({\inimode}_H^\dagger {\inimode}_V^\dagger)/6\nonumber\\
&+&|z_{\mathrm{w}}|^5e^{i5\phi_{\mathrm{w}}}{w_j^\dagger}^5/120.
\end{eqnarray}
Only the first term is necessary to observe the state $\ket{\Delta_5}$. The other terms significantly modify the desired state.

\begin{figure}[thp]
\includegraphics{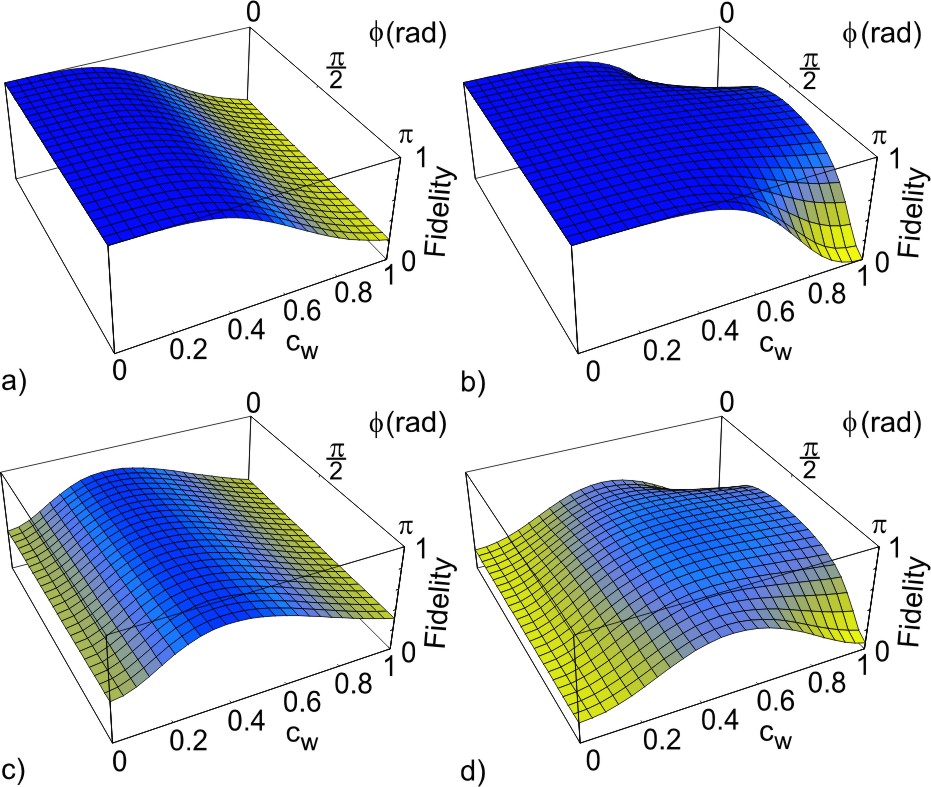}
\caption{\label{fig-simfid} (Color online) Calculated fidelity of $W_4$ (a) and ${GHZ}_4^+$ (b) states when considering only five photon contributions (\eref{eq-5phot}). Calculated fidelity of $W_4$ (c) and ${GHZ}_4^+$ (d) states when considering also six photon contributions. All calculations assume a strength of the SPDC source of $|z_{\mathrm{dc}}|=0.17$, which we experimentally observed for our set-up.}
\end{figure}

Exemplarily, we calculate the fidelity to the ideal $GHZ_4^+$ and $W_4$ states when the photons from \eref{eq-5phot} are symmetrically distributed onto five spatial modes and the respective projective measurement is performed. We obtain
\begin{eqnarray*}
F_{W_4}&=&1/\big[1+|z_{\mathrm{w}}|^4/(9|z_{\mathrm{dc}}|^2)\big],\\
F_{GHZ_4^+}&=&1-\frac{1}{2+36\frac{|z_{\mathrm{dc}}|^2}{|z_{\mathrm{w}}|^4}-12\frac{|z_{\mathrm{dc}}|}{|z_{\mathrm{w}}|^2}\cos{(2\phi_{\mathrm{w}})}},
\end{eqnarray*}
see \fref{fig-simfid}(a) and (b). Both fidelities are better than $>0.99$ for $|z_{\mathrm{w}}|<0.2$, whereas for higher $|z_{\mathrm{w}}|$ the fidelity decreases rapidly. This is the case as with increasing $|z_{\mathrm{w}}|$ the second term of \eref{eq-5phot} grows relatively stronger than the first term and, thus, spoils the state quality. Obviously the relative phase $\phi_{\mathrm{w}}$ becomes only relevant for $F_{GHZ_4^+}$. There, the highest fidelity values are found for $\phi_{\mathrm{w}}=\pi/2$. 

A second effect causes the admixture of correlated noise, which reduces the fidelity. This admixture is produced by the detection of additional five-fold coincidences that originate from six or more photons (higher order emissions from both, the SPDC and the WCB), where multiple photons are registered by the same detector or some photons are not registered at all. As this leads to additional noise, the quality of the observed states is dependent on the photon detection efficiency. For our set-up we determined an efficiency for the photon coupling to the single mode fiber of $\eta_\mathrm{c}\approx\frac{1}{3}$ and a detection efficiency of $\eta_\mathrm{d}\approx\frac{1}{3}$. For calculating the fidelity in that case these loss channels are accounted for by additional beam splitters with ancillary output modes \cite{Kok00}, where reflected photons are lost, and transmitted photons (with probability $\eta_i$) correspond to detectable photons. We consider for this calculation all photon terms of five photons (see \eref{eq-5phot}) and the next higher order contribution from six photons, which are obtained from the multiplication of \eref{eq-SPDC} with \eref{eq-wcb}. The numerical results are shown in \fref{fig-simfid}(c) and (d). The fidelity of the state $\ket{W_4}$ reaches its maximum of $0.776$ for $|z_{\mathrm{w}}|=0.39$ independent of $\phi_{\mathrm{w}}$. For larger $|z_{\mathrm{w}}|$ the fidelity decreases due to the increase of the multiple photon terms of the WCB. For lower $|z_{\mathrm{w}}|$ the fidelity decreases as the contribution from the third order SPDC emission constitutes the major source of noise. The fidelity of the state $\ket{GHZ_4^+}$ reaches its maximum of $0.701$ for $|z_{\mathrm{w}}|=0.6$. Again, it is phase dependent with maximal values for $\phi_{\mathrm{w}}=\pi/2$. The dependence on $|z_{\mathrm{w}}|$ follows the same arguments as given for the $W_4$ state. 

The calculations show that the fidelity of each state is still high enough to demonstrate, e.g., four-photon entanglement via an entanglement witness, as for the state $\ket{W_4}$ ($\ket{GHZ_4^+}$) a fidelity larger than 0.75 (0.5) is sufficient for this purpose \cite{Haf05}(\cite{Tot05pra}). However, in the considerations so far we neglected other experimental imperfections, like spectral distinguishability of photons. For pulsed type II SPDC it is known that the broad pump spectrum results in the generation of photons with partial spectral distinguishability \cite{Gri97,Kel97}, which leads to an additional reduction in state fidelity. For example, the state $\ket{D_4^{(2)}}$ described in \sref{sec-2a} was observed with a fidelity of $F_{\mathrm{exp}}=0.844\pm0.008$ \cite{Kie07}. This fidelity value can be partly ascribed to higher order contributions of the SPDC emission, which give a reduction in fidelity of about $9\%$ \cite{Kie07PHD}. However, the missing $7\%$ can be attributed to a remaining degree of distinguishability of the SPDC photons and non-ideal optical components. It is reasonable to expect that at least the same additional reduction of the fidelity in the proposed implementation occurs. Then, the fidelity for the state $\ket{W_4}$ is below the threshold for proving four-partite entanglement directly.

For this reason we suggest to use a (heralded) single photon source instead of a WCB in order to achieve higher fidelity values. The utilization of a single photon source avoids on the one hand noise from higher order contributions of the WCB and therefore also phase-dependence of the state. On the other hand, even noise from the SPDC alone would become negligible as the heralding signal from the single photon source serves as a trigger for a valid detection event, and thus, the SPDC noise is suppressed. Alternatively, one could realize the state $\ket{D_6^{(3)}}$ with photons coming from the third order SPDC emission, however, on the cost of introducing again SPDC higher order noise. Both alternative implementations demand new and stronger photon sources, which are currently being developed.

\section{Conclusion}

We have demonstrated possibilities for the observation of SLOCC-inequivalent families of three- and four-qubit entangled states. They are based on the property of the states $\ket{D_4^{(2)}}$ and $\ket{\Delta_5}$ to allow access to different classes of quantum states via projective measurements on single qubits.

We experimentally demonstrated that indeed all types of three-qubit entangled states can be obtained from $\ket{D_4^{(2)}}$. We presented a scheme how the states $\ket{\Delta_5}$ can be observed experimentally. As this requires the use of an additional photon, it still poses a considerable challenge when reasonable count rates are to be achieved. We could demonstrate that the most simple approach, i.e., substituting the single photon with a weak coherent beam, leads to a drastic reduction of the state quality. Yet, we identified two alternative possibilities to realize the powerful scheme we presented, which both seem achievable in the near future. 

Altogether our scheme is an alternative method \cite{Wie08,Bas07qph} for the observation of many different multi-partite entangled states. We are optimistic that sources for the observation of the presented states will soon be available and that schemes relying on the same kind of approach will allow the observation of many other interesting quantum states in the future. 

\begin{acknowledgments}
We would like to thank Enrique Solano for stimulating discussions. We acknowledge the support of this work by the DFG-Cluster of Excellence MAP, the EU Project QAP and the DAAD exchange program. W.W.~acknowledges  support by QCCC of the Elite Network of Bavaria and the Studienstiftung des dt.~Volkes.

\end{acknowledgments}


\vspace{\baselineskip}
\copyright \, 2009 The American Physical Society


\begin{thebibliography}{10}

\bibitem{Ben00}
C.~H. Bennett and D.~P. DiVincenzo,
\newblock Nature {\bf 404}, 247 (2000).

\bibitem{Dur00}
W.~D\"ur, G.~Vidal, and J.~I. Cirac,
\newblock Phys. Rev. A {\bf 62}, 062314 (2000).

\bibitem{Ver02}
F.~Verstraete, J.~Dehaene, B.~DeMoor, and H.~Verschelde,
\newblock Phys. Rev. A {\bf 65}, 052112 (2002).

\bibitem{Lam06}
L.~Lamata, J.~Leon, D.~Salgado, and E.~Solano,
\newblock Phys. Rev. A {\bf 74}, 052336 (2006).

\bibitem{Bou99}
D.~Bouwmeester, J.-W. Pan, M.~Daniell, H.~Weinfurter, and A.~Zeilinger,
\newblock Phys. Rev. Lett. {\bf 82}, 1345 (1999);
J.-W. Pan, M.~Daniell, S.~Gasparoni, G.~Weihs, and A.~Zeilinger,
\newblock Phys. Rev. Lett. {\bf 86}, 4435 (2001);
M.~Eibl {\em et~al.},
\newblock Phys. Rev. Lett. {\bf 90}, 200403 (2003);
M.~Eibl, N.~Kiesel, M.~Bourennane, C.~Kurtsiefer, and H.~Weinfurter,
\newblock Phys. Rev. Lett. {\bf 92}, 077901 (2004);
Z.~Zhao {\em et~al.},
\newblock Nature {\bf 430}, 54 (2004);
H.~Mikami, Y.~Li, K.~Fukuoka, and T.~Kobayashi,
\newblock Phys. Rev. Lett. {\bf 95}, 150404 (2005);
N.~Kiesel {\em et~al.},
\newblock Phys. Rev. Lett. {\bf 95}, 210502 (2005);
P.~Walther {\em et~al.},
\newblock Nature {\bf 434}, 169 (2005);
C.-Y. Lu {\em et~al.},
\newblock Nature Physics {\bf 3}, 91  (2007).

\bibitem{Wie08}
W.~Wieczorek {\em et~al.},
\newblock Phys. Rev. Lett. {\bf 101}, 010503 (2008).

\bibitem{Cab99}
C.~Cabrillo, J.~I. Cirac, P.~Garc\'ia-Fern\'andez, and P.~Zoller,
\newblock Phys. Rev. A {\bf 59}, 1025 (1999);
S.~Bose, P.~L. Knight, M.~B. Plenio, and V.~Vedral,
\newblock Phys. Rev. Lett. {\bf 83}, 5158 (1999);
C.~Skornia, J.~vonZanthier, G.~S. Agarwal, E.~Werner, and H.~Walther,
\newblock Phys. Rev. A {\bf 64}, 063801 (2001);
L.-M. Duan, M.~D. Lukin, J.~I. Cirac, and P.~Zoller,
\newblock Nature {\bf 414}, 413 (2001);
L.-M. Duan and H.~J. Kimble,
\newblock Phys. Rev. Lett. {\bf 90}, 253601 (2003);
C.~Simon and W.~T.~M. Irvine,
\newblock Phys. Rev. Lett. {\bf 91}, 110405 (2003);
C.~Thiel, J.~von Zanthier, T.~Bastin, E.~Solano, and G.~S. Agarwal,
\newblock Phys. Rev. Lett. {\bf 99}, 193602 (2007).

\bibitem{Bas07qph}
T.~Bastin {\em et~al.},
\newblock arXiv:0710.3720 [quant-ph]  (2007).

\bibitem{Dic54}
R.~H. Dicke,
\newblock Phys. Rev. {\bf 93}, 99 (1954).

\bibitem{Kie07}
N.~Kiesel, C.~Schmid, G.~Toth, E.~Solano, and H.~Weinfurter,
\newblock Phys. Rev. Lett. {\bf 98}, 063604 (2007).

\bibitem{Sch07ProcNato}
C.~Schmid {\em et~al.},
\newblock The entanglement of the symmetric four-photon dicke state,
\newblock in {\em Quantum Communication and Security}, edited by M.~Zukowski,
  S.~Kilin, and J.~Kowalik, p. 113, IOS Press (ISBN 978-1-58603-749-9),
  Netherlands, 2007.

\bibitem{ScP07}
C.~Schmid, N.~Kiesel, W.~Wieczorek, R.~Pohlner, and H.~Weinfurter,
\newblock SPIE Proc. 6780 , 67800E (2007).

\bibitem{Lam07}
L.~Lamata, J.~Leon, D.~Salgado, and E.~Solano,
\newblock Phys. Rev. A {\bf 75}, 022318 (2007).

\bibitem{Cof00}
V.~Coffman, J.~Kundu, and W.~K. Wootters,
\newblock Phys. Rev. A {\bf 61}, 052306 (2000).

\bibitem{Sen03a}
A.~Sen(De), U.~Sen, and M.~\.Zukowski,
\newblock Phys. Rev. A {\bf 68}, 032309 (2003).

\bibitem{StatsTrafo}
In order to obtain the family of states
  $\mu(\alpha,\markproj{\alpha},\epsilon,\markproj{\epsilon})\ket{GHZ_4}$
  $+\nu(\alpha,\markproj{\alpha},\epsilon,\markproj{\epsilon})\ket{D_4^{(2)}}$
  we need to apply the local transformations $(T_2T_1)^{\otimes 4}$ on the
  states of \eref{eq-4photext}, where
  $T_{1}=\left(\begin{array}{cc}x&y\\x&-y\end{array}\right)$ and
  $T_2=(z\openone-\sigma_z)^{0.25}$ with
  $x=\exp{(i\epsilon/2)}\sqrt{\markproj{\alpha}\beta/2}$,
  $y=\exp{(-i\markproj{\epsilon}/2)}\sqrt{\alpha\markproj{\beta}/2}$ and
  $z=0.75[\exp{(i(\markproj{\epsilon}-\epsilon)/2)}\alpha\markproj{\alpha}$
  $+\exp{(-i(\markproj{\epsilon}-\epsilon)/2)}\beta\markproj{\beta}]/$
  $\sqrt{\alpha\markproj{\alpha}\beta\markproj{\beta}}$. A simple calculation
  shows that the absolute value of $\mu$ (and accordingly $\nu$) can take all
  values between $0$ and $1$. For particular parameters other types of states
  can be obtained, e.g., if only the local transformations $T_1^{\otimes 4}$
  are applied on the states of \eref{eq-4photext} we can obtain for example the
  state $\sqrt{6/7}\ket{HHHH}-\sqrt{1/7}\ket{D_4^{2}}$
  [$\alpha=\markproj{\alpha}$ $=1/\sqrt{2}$ and
  $\epsilon=2\arccos{(2/3)}+\markproj{\epsilon}$].

\bibitem{Bas08tbd}
T.~Bastin~{\emph{et al.}},
\newblock in preparation  (2008).

\bibitem{Kok00}
P.~Kok and S.~L. Braunstein,
\newblock Phys. Rev. A {\bf 61}, 042304 (2000).

\bibitem{Wei01}
H.~Weinfurter and M.~\.Zukowski,
\newblock Phys. Rev. A {\bf 64}, 010102(R) (2001).

\bibitem{Bou04}
M.~Bourennane {\em et~al.},
\newblock Phys. Rev. Lett. {\bf 92}, 087902 (2004).

\bibitem{Sch08DS}
C.~Schmid {\em et~al.},
\newblock Phys. Rev. Lett. {\bf 100}, 200407 (2008).

\bibitem{Aci01}
A.~Acin, D.~Bru{\ss}, M.~Lewenstein, and A.~Sanpera,
\newblock Phys. Rev. Lett. {\bf 87}, 040401 (2001).

\bibitem{Haf05}
H.~H\"affner {\em et~al.},
\newblock Nature {\bf 438}, 643 (2005).

\bibitem{Oxb05}
M.~Oxborrow and A.~G. Sinclair,
\newblock Cont. Phys. {\bf 46}, 173 (2005).

\bibitem{LoS05}
B.~Lounis and M.~Orrit,
\newblock Rep. Prog. Phys. {\bf 68}, 1129 (2005).

\bibitem{Roh05}
P.~P. Rohde, T.~C. Ralph, and M.~A. Nielsen,
\newblock Phys. Rev. A {\bf 72}, 052332 (2005).

\bibitem{Rar99}
J.~G. Rarity and P.~R. Tapster,
\newblock Phys. Rev. A {\bf 59}, R35 (1999).

\bibitem{Man95}
L.~Mandel and E.~Wolf,
\newblock {\em Optical Coherence and Quantum Optics} (Cambridge University
  Press, 1995).

\bibitem{Lam02}
A.~Lamas-Linares, C.~Simon, J.~C. Howell, and D.~Bouwmeester,
\newblock Science {\bf 296}, 712 (2002).

\bibitem{Sci05}
F.~Sciarrino and F.~DeMartini,
\newblock Phys. Rev. A {\bf 72}, 062313 (2005).

\bibitem{Mar05}
F.~DeMartini, F.~Sciarrino, and V.~Secondi,
\newblock Phys. Rev. Lett. {\bf 95}, 240401 (2005).

\bibitem{Sun07}
F.~W. Sun {\em et~al.},
\newblock Phys. Rev. Lett. {\bf 99}, 043601 (2007).

\bibitem{Tot05pra}
G.~T\'oth and O.~G\"uhne,
\newblock Phys. Rev. A {\bf 72}, 022340 (2005).

\bibitem{Gri97}
W.~P. Grice and I.~A. Walmsley,
\newblock Phys. Rev. A {\bf 56}, 1627 (1997).

\bibitem{Kel97}
T.~E. Keller and M.~H. Rubin,
\newblock Phys. Rev. A {\bf 56}, 1534 (1997).

\bibitem{Kie07PHD}
N.~Kiesel,
\newblock {\em Experiments on Multiphoton Entanglement},
\newblock PhD thesis, Ludwig-Maximilians-Universit\"at M\"unchen, 2007.

\end{thebibliography}
\end{document}